\documentstyle[prl,aps,multicol,epsf]{revtex}
 
\begin{document}

\title{The no-slip condition for a mixture of two liquids}
\author{Joel Koplik}
\address{Benjamin Levich Institute and Department of Physics \\
City College of the City University of New York, New York, NY 10031}
\author{Jayanth R. Banavar}
\address{Department of Physics and Center for Materials Physics \\
Pennsylvania State University, University Park, PA 16802}

\date{\today}
\maketitle

\begin{abstract}
When a mixture of two viscous liquids flows past a solid wall
there is an ambiguity in the use of the no-slip boundary condition.
It is not obvious whether the mass-averaged velocity, the
volume-averaged velocity, the individual species velocities, all or none of 
the above, or none of the above should exhibit no-slip.  
Extensive molecular dynamics simulations of the Poiseuille flow of 
mixtures of coexisting liquid species past an atomistic wall indicate
that the velocity of each individual liquid species satisfies the 
no-slip condition and, therefore, so do mass and volume averages.

\end{abstract}

\begin{multicols}{2}
The no-slip boundary condition in fluid mechanics states that 
the velocity of a fluid at a solid wall equals the velocity of the wall.  This
condition has a lengthy history \cite{goldstein} and solid experimental 
support for the case of Newtonian liquids, but its microscopic basis has 
been understood only relatively recently based on molecular dynamics 
simulations \cite{alder,hlc,kbw,tr}.  The validity of the no-slip condition is
far from automatic, since for example gases \cite{gas} and non-Newtonian 
liquids \cite{nn} are known to exhibit slip, and in this paper we consider the
question of the appropriate wall boundary condition for a {\em mixture} of two
liquids.  The question is relevant to any flow problem involving miscible
liquids, such as double diffusive convection, where the composition is 
spatially varying.  

Since there is no satisfactory analytic argument for the origin of no-slip
even in a one-component fluid, it is not feasible to obtain a theoretical 
derivation here
from first principles.  Likewise, it is difficult to address the question
experimentally because the absence of an accepted and complete theory for 
fluid mixtures prevents one from, for example, extracting a slip length by
measuring flux {\it vs}.\ pressure drop in Poiseuille flow.  In recent work 
it is common to assume that the mass-averaged velocity satisfies no slip;  
see, {\it e.g.,} Perera and Sekerka \cite{perera}, and Liao and Joseph 
\cite{liao}.  Aside
from some theoretical justification by Camacho and Brenner \cite{camacho},
and some supporting molecular dynamics simulations by Mo and Rosenberger 
\cite{mo} for gases, there is some intuitive plausibility for focusing on 
mass-averaged velocity because in the microscopic interactions between fluid 
and wall atoms, the momentum is the relevant dynamical variable.  On the other
hand, one might say that local volume averaging is an appropriate scheme to
treat mixture problems, and boundary conditions should involve this variable.
Alternatively, one might be influenced by the kinetic theory arguments of
Maxwell \cite{maxwell} for slip in gases, and consider the momentum exchange 
in inter-species collisions \cite{hinch}, and conclude that the mass ratio
will enter in the boundary condition.  Yet another line of reasoning is to
assert that a fluid molecule moves in a potential due to the nearly-fixed wall
atoms and the nearly-randomly fluctuating neighboring fluid molecules, in 
which there is little difference between a homogeneous liquid and a mixture,
and therefore the individual species velocities should satisfy no-slip.
Unfortunately, none of these heuristic arguments is completely decisive in 
itself.

In order to provide an unbiased and fundamental calculation of the wall
boundary condition for a liquid mixture, we have conducted systematic 
molecular dynamics (MD) calculations \cite{at} of the Poiseuille flow of 
systems of several thousand liquid atoms in a channel made of atomistic 
solid walls, in cases with or without a concentration gradient.
Previous MD studies of this nature (see, \cite{arfm} for a general review)
indicate that the continuum aspects of fluid flow are well reproduced in this
way, and we have been able to obtain unambiguous numerical results for the
showing that the individual species velocities satisfy the no-slip condition.
We emphasize that this statement holds for {\em liquids};  in a mixture of
gases one has the more difficult question of studying the variation of 
a non-zero slip length with the properties of the components of the mixture.

The mechanics of the MD simulations are very similar to those used by us 
in a study of the sliding plate problem \cite{slide}. We first consider the
channel flow of a {\em homogeneous} Newtonian liquid made of two types of
atom.  The simulated system consists of 8000 fluid atoms
flowing between two solid plane atomic walls.  The atoms
interact with each other via two-body Lennard-Jones (LJ) potentials, 
$ V_{ij}(r) = 4\epsilon\left[ ({r/\sigma})^{-12}-c_{ij}
({r/\sigma})^{-6} \right]$,
cutoff at separation $r=2.5$, where $c_{ij}$ ($i,j$=1,2,w) is an adjustable 
inter-species interaction strength.  Hereafter, all quantities in this paper 
are non-dimensionalized using the energy scale $\epsilon$, the length scale
$\sigma$, and a reference mass $m$, giving a natural timescale
$\tau=\sigma(m/\epsilon)^{1/2}$.  The walls are tethered to fixed
lattice sites by harmonic springs \cite{tr}, and otherwise interact with all
other wall and fluid atoms via an LJ potential with an appropriate $c_{ij}$, 
allowing them to exchange energy and momentum with the fluid.  The system is
three dimensional, with the walls parallel to the $x$-$z$ plane separated by
a gap of width 18$\sigma$ in the $y$-direction, periodic boundaries in $x$ and
$z$, and average flow in the $x$-direction produced by applying a
gravitation-like force $m_ig\hat{x}$ to each fluid atom, with $g=0.01$.  
In most cases, we 
extract the heat generated by viscous friction by a local Nos\'e-Hoover 
thermostat \cite{nh} which couples each atom to a heat bath, providing a 
constant temperature system.

\begin{figure}[h]
\epsfxsize=8.0cm
\epsffile{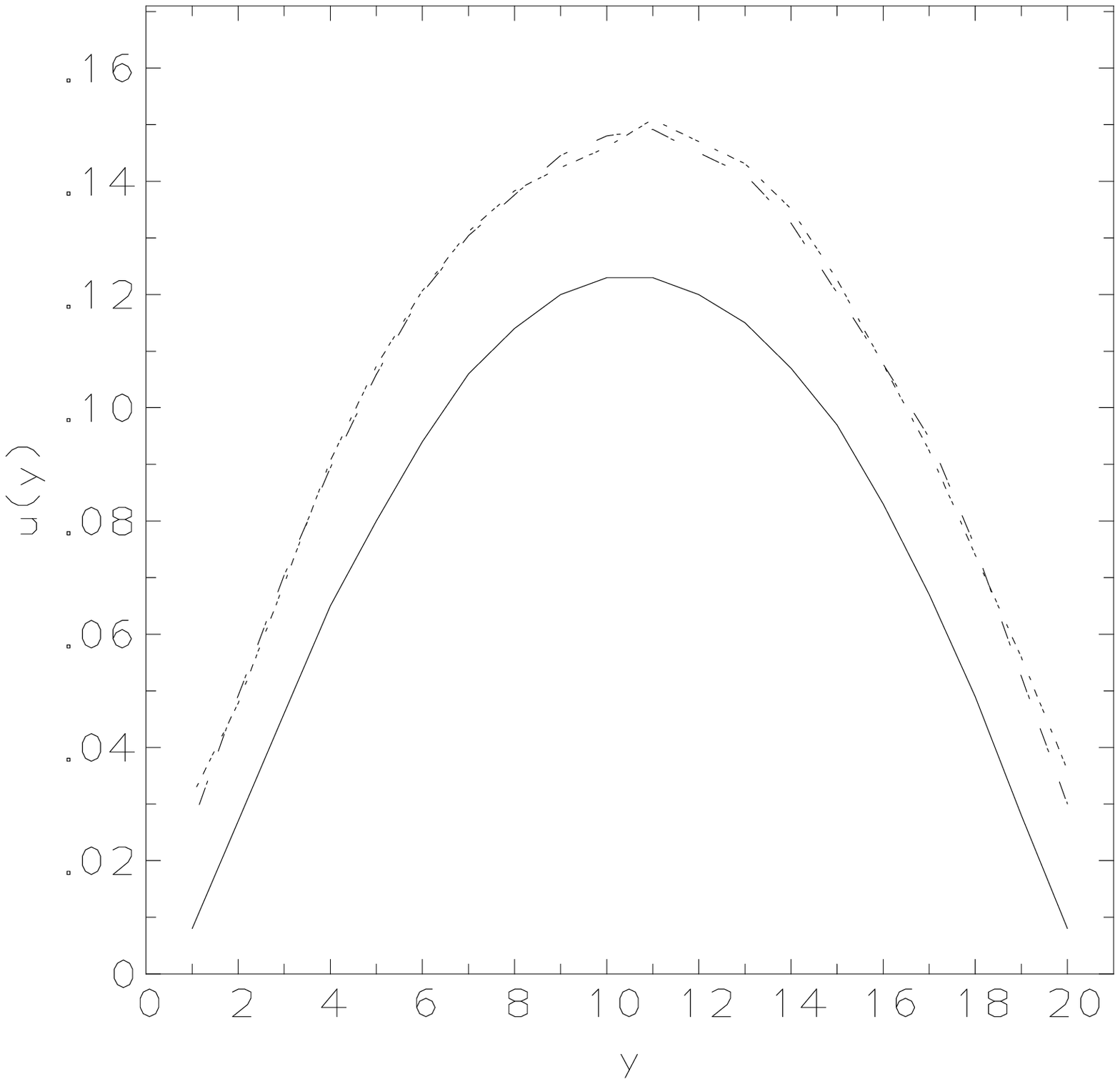}
\vspace{0.1in}
\epsfxsize=8.0cm
\epsffile{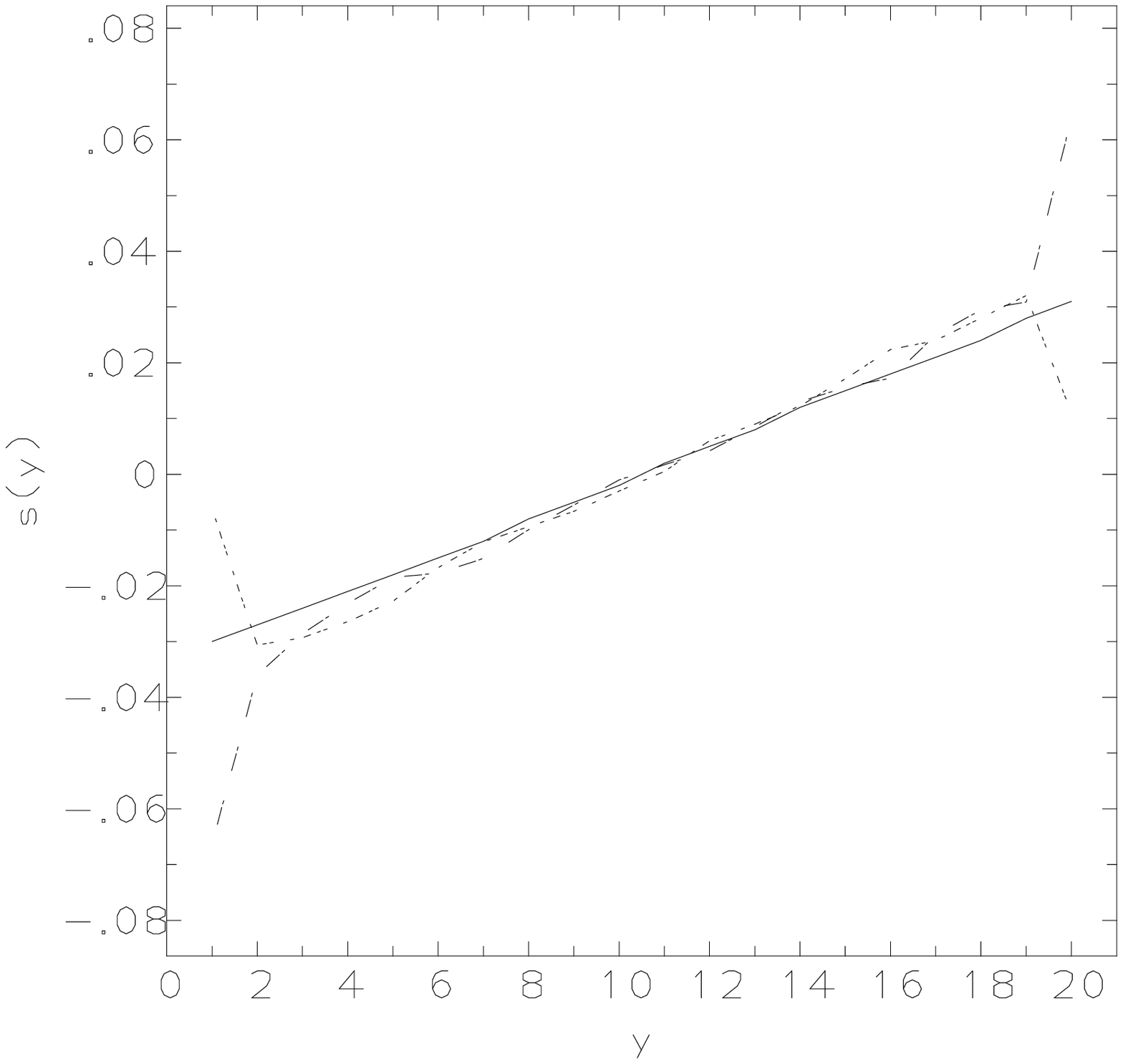}
\vspace{0.1in}
\narrowtext
\caption{(a) Velocity and (b) shear stress profiles for a single liquid
(solid line), and for the individual species of a uniform mixture (dashed
and dotted lines.} 
\end{figure}

In Fig.~1, we show typical results for the tangential velocity $u(y)$ and 
the shear stress $s(y)$.  The three sets of curves are a reference 
one-species system (unit mass and unit fluid-wall 
coupling) plotted as a solid line, and then the two distinct species for 
the mixture case shown as dashed or dotted lines.  In this case there are 
an equal number of atoms of each species, and their parameters  
have been chosen as $m_1=1.5$, $m_2=0.75$, $c_{12}=0.9$, $c_{1w}=0.75$ 
and $c_{2w}=0.25$.
The $y$-coordinate is such that $y$=0 and 21 are the centers of the 
two innermost layers of wall atoms confining the fluid, and there are 
20 sampling bins
for the fluid velocity and stress, centered at integer values of $y$.
All velocity profiles are parabolas which go to zero in the vicinity of the
wall.  The pure liquid has a stronger wall attraction, so its
extrapolated velocity vanishes slightly further away from the wall.  In 
other mixture simulations of this type, if $c_{iw}=1$ the species velocities
vanish at approximately the same position as the pure liquid, while 
when $c_{iw}>1$ the species velocities vanish further away.
The difference in the height of the parabola of pure liquid compared to the
mixture arises from the latter having a different effective viscosity.  
The stress profiles are linear in the interior of the channel, 
as expected in Poiseuille flow, but the component of the mixture which is more 
(less) strongly attracted to the wall has a higher (lower) stress value there,  
accompanied by a sharp increase (decrease) in density.
We have repeated this calculation for a number of uniformly mixed systems with
different parameter values, and in all cases the two components have 
essentially the same velocity profile.  When the mixture concentration is
50\%, the shear stress profiles agree, but for other concentrations we observe
distinct slopes: the component of higher concentration has a steeper stress
gradient, and a correspondingly higher effective viscosity, and a higher 
pressure as well.  In one test case we omitted the thermostat and
allowed the system's temperature to rise.  The result is again a parabolic
velocity profile which goes to zero near the walls, but with larger numerical 
values reflecting the decrease of viscosity with temperature.

Although we do not have a precise mathematical model to elucidate the numerical
observations, the physics is reasonably clear.  The molecular origin of no-slip
is that fluid molecules near the wall are on the one hand dragged along in
the direction of the net flow by their neighbors further away from the wall, 
and on the other hand are pushed up against the wall by the crowding of
molecules in a dense liquid, where they interact with the effectively 
corrugated potential of the wall.  While the molecules near the wall are not
literally stuck there, their translation speed is much reduced as they interact
with the almost-fixed wall atoms, and a typical trajectory \cite{kbw} is a
random walk with a drift in the direction of the applied force and some
transient localization near the wall.  When the atoms are sorted by sampling
bins (or averaged over position by the finite spatial resolution of any 
realistic laboratory measurement) the average velocity decreases systematically
as the wall is approached, and tends to zero near the wall.  The precise
value of ``near'' depends on the details of the wall-fluid interaction, but for
reasonable choices, the distance from the nominal wall position is at most one
or two atomic sizes, and zero from a macroscopic viewpoint.  The key point 
about a mixture of liquids is that the
qualitative argument just given does not depend in any way on the details 
of the interactions between the fluid atoms or whether there is more than one
fluid species.  Thus, in a dense liquid in good contact with the wall, any
species should satisfy the no-slip condition.  (The qualification concerning
the wall contact is to distinguish Newtonian from non-Newtonian fluids or
suspensions:  in a polymer melt only part of a molecule adjacent to a wall need
be in atomic contact, while in the case of a polymer solution or suspension 
there can be a depletion region near the wall.)

A more interesting case arises when the relative concentration of the species
varies with position along the wall.  The intuitive argument just presented
is unchanged by this complication, but the numerical verification is rather
more involved.  If the variation is weak, one may treat a local
region by the homogeneous simulation just described, using the appropriate 
local concentration value.  More generally, we wish to consider 
a rapid linear concentration gradient along the direction channel;  it 
is trivial to initialize the system in this
way, but an additional ingredient is needed to prevent the gradient from 
diffusing away with time.  To this end, we divide the channel along the 
($x$) flow direction into 40 ``concentration bins'', where in bin 1 
(21) the concentration
of species 1 is fixed at 0.75 (0.25) at all time.  These values are maintained
by changing the identity of atoms chosen at random from species 1 to 2, or vice
versa, until the concentration has the desired value.  In these two bins 
a kind of Maxwell demon is operating, and the fluid and flow properties 
there may not be fully realistic, but the other 38 bins are untouched and 
presumably faithful models of two-fluid coexistence at whatever local value 
of concentration occurs.  A more sophisticated computation \cite{vanswol}
would fix the concentration in the two reference bins by coupling them to
particle reservoirs using a grand canonical
Monte Carlo procedure, which has realistic fluid behavior
everywhere, but the simpler method suffices here.  

In the absence of net flow
the procedure just described indeed produces a sawtooth concentration profile
(Fig.~2, dashed line) but when motion occurs the reservoir values of 
concentration are convected downstream and the result is a profile which is
not piecewise linear (Fig.~2, solid line).  In fact, as one might expect,
the observed profiles are similar to solutions of the convection-diffusion 
equation.  One then has a numerical dilemma:  
in order to have a statistically significant velocity profile, it is 
preferable to have a strong forcing acceleration and large values for the mean
velocity, but in this case the concentration profile is unsuitable.
In Lennard-Jones systems a liquid freezes if the temperature is
too low, corresponding to an O(1) minimum liquid thermal velocity,
and if the mean velocities are much below this value they are obscured by 
thermal noise.  However, if the mean velocities are too large, the resulting
concentration profile approaches a step, with a very narrow transition region,
and within each step the species concentrations are constant.  

\begin{figure}[h]
\epsfxsize=8.0cm
\epsffile{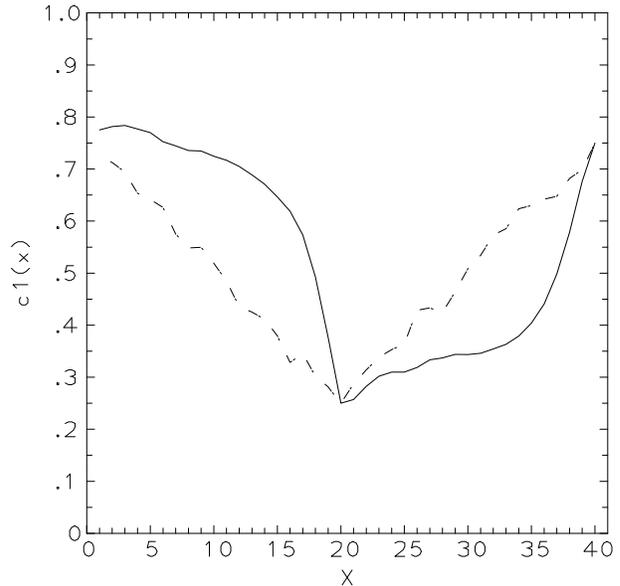}
\vspace{0.1in}
\narrowtext
\caption{ Concentration profiles for one species in a mixture at rest (dashed 
line) and under Poiseuille flow (solid line).The first curve is noisier
because it results from a much shorter averaging interval.} 
\end{figure}

The profile shown in Fig.~2 was obtained with a 
rather weak forcing, $g=0.001$, and in order to observe a meaningful velocity 
profile we have simply averaged over a very long time interval of 2500$\tau$ 
(and also doubled the length of the system so as to 
have 16000 fluid atoms).  The
resulting computation required about 2 months on an HP-735/125 workstation,
and gave the concentration profile above and the subsequent velocity 
plots, which are still somewhat noisy, but clearly display no-slip 
velocity profiles.

In Fig.~3, we show some typical fluid 1 velocity profiles, $u_1(x,y)$, 
as a function of $y$ for selected $x$-bins.  The concentration at each $x$
may be found from the solid curve in Fig.~2.  Each curve is roughly parabolic
and tends to zero near the wall positions.  Because each point involves a 
average of only 5-15 atoms, compared to 200 in the uniform mixture case in
Fig.~1, the statistical fluctuations are much stronger.  If these curves are
< averaged, one finds a smooth parabola resembling Fig.~1a.  The profiles
of the second fluid and the other half of the channel are qualitatively the
same.  Note that there is no qualitative difference as a function of
concentration, nor between bin 1, where the atoms' identity was varied 
unphysically, and the other bins.  This insensitivity is analogous to that
found as the method of thermostatting varies, and is characteristic of MD 
simulations of this type.  

\begin{figure}[h]
\epsfxsize=8.0cm
\epsffile{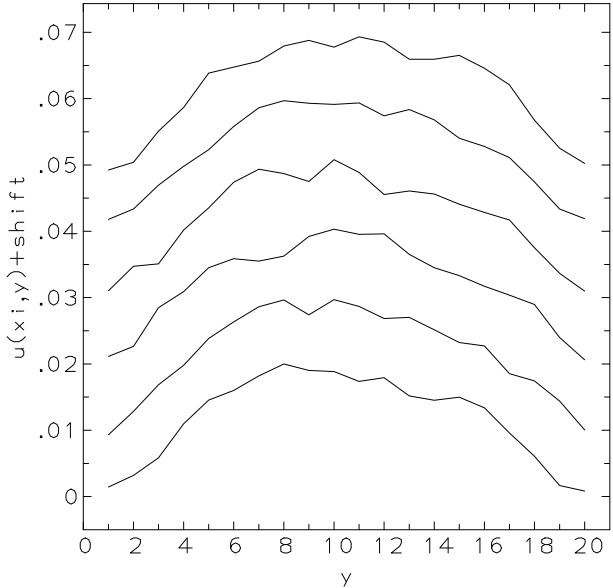}
\vspace{0.1in}
\narrowtext
\caption{ Velocity profiles of species 1 as a function of $y$ for bins $x_i= 
1,4,12,16,20$, corresponding to the concentration gradient in Fig.~2.
The successive curves have been shifted upwards by 0.01 for clarity.}
\end{figure}

We have shown that in a mixture of liquids the species velocities satisfy the
no-slip condition and, {\em a fortiori}, so do mass and volume averages.
The details of liquid-solid interactions may affect the behavior within atomic
distances of the walls, but macroscopically the classical boundary condition
holds.  Although we have not pursued the matter here, this type of data and 
calculation can be used to test detailed theories of the dynamics of 
liquid mixtures.  Channel flows of uniform mixtures are computationally
efficient, although in concentration gradients the results 
can be noisy and time-consuming to obtain.  However, molecular dynamics
simulations can again provide otherwise inaccessible information.
 
\smallskip
We thank E. J. Hinch, R. Jackson, R. F. Sekerka and M. G. Worster for
stimulating discussions.  This research was supported by the NASA Microgravity
Science and Applications Division.

\end{multicols}

\end{document}